\documentstyle[aps,preprint]{revtex}

\begin{document}
\draft
\title{ Dynamics of fermions coupling to a $U(1)$ gauge
field in the limit $e^2\rightarrow\infty$
}
\author{Tai-Kai Ng}
\address{
Department of Physics,
Hong Kong University of Science and Technology,\\
Clear Water Bay Road,
Kowloon, Hong Kong
}
\date{ \today }
\maketitle
\begin{abstract}
   We study in this paper the properties of a gas of fermions coupling
to a $U(1)$ gauge field at wavevectors $q<\Lambda<<k_F$ at dimensions 
larger than one, where $\Lambda<<k_F$ is a high momentum cutoff and 
$k_F$ is the fermi wave vector. In particular, we shall consider the
$e^2\rightarrow\infty$ limit where charge and current fluctuations at 
wave vectors $q<\Lambda$ are forbidden. Within a bosonization approximation, 
effective actions describing the low energy physics of the system are
constructed, where we show that the system can be described as a fermion 
liquid formed by chargeless quasi-particles which has vanishing 
wavefunction overlap with the bare fermions in the system.

\end{abstract}

\pacs{PACS Numbers: 71.27.+a, 74.25.-q, 11.15.-q }

\narrowtext

\section{introduction}
    In the last few years, there has been enormous interests in the
study of $U(1)$ gauge theories of fermionic systems in dimensions
higher than one, as a result of appearance of effective gauge
theories in the $t-J$ type models\cite{t1} and also in the studies of the
$\nu={1\over2}$ Fractional Quantum Hall state\cite{t2,hal}. In 
particular, the behaviour of the systems in the strong-coupling regime 
where charge and current fluctuations are forbidden (confined) are of 
special interests. In a previous paper we studied the effects of a
longitudinal gauge field (Coulomb interaction) on a system of spinless
fermions in the limit $e^2\rightarrow\infty$. Within a bosonization
approximation, we found that the ground state and low energy 
excitations of the system are described by a fermi liquid with 
chargeless quasi-particles\cite{ng1}. In this paper we shall generalize 
our approach\cite{ng1} to consider the $e^2\rightarrow\infty$ limit of 
a system of spin $S=1/2$, charge $e$ fermions minimally coupled to a 
$U(1)$ gauge field $(A_0,\vec{A})$ in dimensions higher than one. The
gauge field dynamics is described by an effective long distance action
$L_{gauge}=F_{\mu\nu}^2$ with high momentum cutoff $\Lambda<<k_F$, 
where $k_F$ is the fermi wavevector. In this limit any non-uniform charge 
and current density fluctuations $<\rho(\vec{q})>\neq0$ and 
$<\vec{j}(\vec{q})>\neq0$ with wave vector $q<\Lambda$ in the system
cost infinite energy and are forbidden. As a result any physical 
state $|\psi>$ that survives in this limit satisfies the constraints
$\rho(\vec{q})|\psi>=0$ and $\vec{j}(\vec{q})|\psi>=0$, where 
$\rho(\vec{q})$ and $\vec{j}(\vec{q})$ are the charge and current 
density operators, respectively.

  In this paper we shall show that by generalizing our previous 
approach\cite{ng1}, the projection to the physical states
$|\psi>$ where density and current fluctuations are forbidden can be 
carried out in the $e^2\rightarrow\infty$ limit in a bosonization
approximation\cite{b0,b1,b2,b3}, and physical (chargeless) 
single-particle operators that commute with both charge and current 
operators of the 
system can be constructed. The method we use is similar in spirit to
the field theory scheme proposed by Shanjar and Murthy\cite{sm} for
the Fractional Quantum Hall effect, though the details are different. 
Within the approximation effective low energy Lagrangians that 
describe the chargeless particle-hole as well as single-particle 
excitations in the system are obtained. The physical sector of
the system is described by a liquid of 
chargeless fermions and corresponds to a 
kind of 'marginal' fermi liquid in the original fermion description. 
The organization of our paper is as follows: in section II we shall 
outline our mathematical formulation of the bosonization procedure where 
an approximate bosonized action of the system describing both spin
and charge particle-hole excitations will be derived. In section III
we shall study in detail the eigenstates and eigen-energies of the 
particle-hole excitation spectrum in the bosonization approximation 
for arbitrary values of coupling constant $e^2$. In particular we 
shall show how the projection to the physical Hilbert space where 
charge and current fluctuations are forbidden is obtained in the
$e^2\rightarrow\infty$ limit. In section IV we shall consider the single
particle excitations where $S=1/2$, chargeless single fermion operators 
that commute with density and current operators are constructed and the
equation of motion for these fermion operators is derived. An effective 
action in terms of these chargeless fermion operators is constructed 
where we shall show that the system forms a marginal fermi liquid. Our 
results will be summarized in section V where some further comments will 
be given. 
 
\section{mathematical formulation}
    We shall work in the Coulomb gauge $\nabla.\vec{A}=0$ where the 
$A_0$ component of the gauge field is integrated out, resulting in 
an instantaneous Coulomb interaction $v(q)=4\pi{e}^2/q^2$ at
$q<\Lambda$ between fermions. The resulting Hamiltonian of the system is, 
\begin{equation}
\label{ham}
H=\sum_{i\sigma}{1\over2m}\left[\vec{p}_{i\sigma}-{e\over{c}}\vec{A}
(\vec{r}_{i\sigma})\right]^2+{1\over2L^d}\sum_{\vec{q}\neq0,q<\Lambda}
v(q)\rho(\vec{q})\rho(-\vec{q})+H_{photon},
\end{equation}
where $\vec{p}_{i\sigma}$ and $\vec{r}_{i\sigma}$ are the momentum and 
position operators of the $i^{th}$ fermion with spin $\sigma$. $\rho(\vec{q})=
\sum_{\vec{k}\sigma}f^+_{\vec{k}+\vec{q}/2\sigma}
f_{\vec{k}-\vec{q}/2\sigma}$ is the density operator for the fermions.
$f(f^+)_{\vec{k}\sigma}$'s are fermion annihilation (creation) operators. 
$L^d$ is the volume of the system and $c$ is the velocity of light. 
\begin{equation}
\label{photon}
H_{photon}=\sum_{\vec{q}\neq0,q<\Lambda,\lambda}
\omega_q\left(a^+_{\lambda}(\vec{q})a_{\lambda}(\vec{q})+{1\over2}\right),
\end{equation}
where $a(a^+)_{\vec{q}\lambda}$'s are photon annihilation (creation) 
operators with wavevector $\vec{q}$ and polarization $\lambda$. 
$\omega_q=cq$. We have set $\hbar=1$ to simplify notation. Notice that 
the Hartree ($\vec{q}=0$) interaction energy does not appear in the 
Hamiltonian as in usual Coulomb gas problem where overall charge 
neutrality is maintained by presence of a uniform background of opposite 
sign charges. After Fourier transforming and in second quantized 
form, we obtain
\begin{mathletters}
\label{f-p}
\begin{eqnarray}
\label{fermion}
\sum_{i\sigma}{1\over2m}\left[\vec{p}_{i\sigma}-{e\over{c}}\vec{A}
(\vec{r}_{i\sigma})\right]^2 & \rightarrow & \sum_{\vec{k}\sigma}
({k^2\over2m})f^+_{\vec{k}\sigma}f_{\vec{k}\sigma}-{e\over{L}^{d/2}}
\sum_{\vec{q}\lambda}\vec{j}_p(\vec{q}).\vec{A}(\vec{q},\lambda)
\\ \nonumber
& & +\sum_{\vec{q}\lambda}({e^2n_0\over2m})\vec{A}(\vec{q},\lambda).
\vec{A}(-\vec{q},\lambda),
\end{eqnarray}
where $\vec{j}_p(\vec{q})=\sum_{\vec{k}\sigma}(\vec{k}/m)
f^+_{\vec{k}+\vec{q}/2\sigma}f_{\vec{k}-\vec{q}/2\sigma}$ is the
paramagnetic current operator, and
\begin{equation}
\vec{A}(\vec{q},\lambda)=\left({2\pi\over\omega_q}\right)^{1\over2}
\vec{\xi}(\vec{q},\lambda)(a_{\lambda}(\vec{q})+a^+_{\lambda}(-\vec{q})),
\end{equation}
\end{mathletters}
where $\vec{\xi}(\vec{q},\lambda)$ is the polarization vector of photons 
with momentum $\vec{q}$ and polarization mode $\lambda$. Note that all 
sums over $\vec{q}$'s are restricted to $\vec{q}\neq0, |\vec{q}|<\Lambda$
in Eq.\ (\ref{f-p}) and in all the following equations. We have also 
replaced the fermion density operator $\rho(\vec{r})$ by its 
expectation value $n_0$ in the diamagnetic term in Eq.\ (\ref{fermion}). 
This approximation can be justified in the $e^2\rightarrow\infty$ limit 
where density fluctuations of fermions are forbidden. We shall discuss 
this more carefully in section IV.

    The difference between our model and usual Coulomb gas problem has to 
be emphasized here. In usual Coulomb gas problem the high momentum cutoff 
$\Lambda$ is taken to be infinity, or satisfies $\Lambda>>k_F$. In this 
limit a Wigner crystal where electrons position are fixed is expected to 
be formed (in 3D) when $e^2$ becomes large because the potential energy 
term becomes dominating at length scale $\sim{k}_F^{-1}$. In our problem 
where $\Lambda<<k_F$, the interaction is effective only at length scale 
$>>$ inter-particle spacing and formation of Wigner crystal is not 
warranted. In particular we found in Ref.[\cite{ng1}] that the ground 
state of the 
system has large degeneracies in the absence of the kinetic energy term 
and the system may remain in a liquid state when kinetic energy
term is switched on. We note that bosonization method is 
a natural tool to study the liquid state of the problem in this
limit\cite{b2,b3,b4}. The bosonization procedure can be formulated most 
easily by introducing the Wigner function operators\cite{ng1} 
$\rho_{\vec{k}\sigma}(\vec{q})=f^+_{\vec{k}+\vec{q}/2\sigma}
f_{\vec{k}-\vec{q}/2\sigma}$. In the path-integral formulation the 
Wigner operators are introduced in the Imaginary-time action of the 
system through Langrange multiplier fields,
\begin{eqnarray}
\label{lag}
S & = & \int^{\beta}_0d\tau\left[\sum_{\vec{k}\sigma}f^+_{\vec{k}\sigma}
(\tau)({\partial\over\partial\tau}+{k^2\over2m}-\mu)f_{\vec{k}\sigma}
(\tau)-\sum_{\vec{k}\sigma,\vec{q}}i\lambda_{\vec{k}\sigma}(\vec{q},
\tau)\left(\rho_{\vec{k}\sigma}(\vec{q},\tau)-
f^+_{\vec{k}+\vec{q}/2\sigma}(\tau)f_{\vec{k}-\vec{q}/2\sigma}(\tau)
\right)\right.  \\   \nonumber
& & +\left.{1\over{L}^d}\sum_{\vec{q},\vec{k},\vec{k}'}
v(q)\rho_{\vec{k}}(\vec{q},\tau)\rho_{\vec{k}'}(-\vec{q},\tau)
-{\sqrt{2}e\over{L}^{d/2}}\sum_{\vec{k},\vec{q}\lambda}({\vec{k}.\vec{A}
(\vec{q},\lambda,\tau)\over{m}})\rho_{\vec{k}}(\vec{q},\tau)\right] 
+ S'_{photon},
\end{eqnarray}
where $\mu$ is the chemical potential, 
\[
S'_{photon}=S_{photon}+\sum_{\vec{q}\lambda}({e^2n_0\over2m})
\vec{A}(\vec{q},\lambda,\tau).\vec{A}(-\vec{q},\lambda,\tau),
\]
where $S_{photon}$ is the pure photon action, $\rho_{\vec{k}}(\vec{q})={1\over\sqrt{2}}\sum_{\sigma}
\rho_{\vec{k}\sigma}(\vec{q})$ are {\em density} Wigner function 
operators, and $\lambda_{\vec{k}\sigma}(\vec{q})$'s are Lagranger multiplier 
fields introduced to enforce the constraint that the Wigner operators 
are given by $\rho_{\vec{k}\sigma}(\vec{q})=f^+_{\vec{k}+\vec{q}/2\sigma}
f_{\vec{k}-\vec{q}/2\sigma}$. In particular, the original 
Hamiltonian \ (\ref{ham}) is recovered once the 
$\lambda_{\vec{k}\sigma}(\vec{q})$ field is integrated out.
For later convenience we shall also introduce {\em spin} Wigner
operators $\sigma_{\vec{k}}(\vec{q})={1\over\sqrt{2}}\sum_{\sigma}
\sigma\rho_{\vec{k}\sigma}(\vec{q})$. 

  The photon action $S'_{photon}$ can be diagonalized easily to
obtain
\begin{mathletters}
\begin{equation}
\label{sphoton}
S'_{photon}=\sum_{\vec{q}\lambda,i\omega_n}(-i\omega_n+\Omega_q)
b^+_{\lambda}(\vec{q},i\omega_n)b_{\lambda}(\vec{q},i\omega_n),
\end{equation}
where $b(b^+)$'s are new photon eigenmodes with eigenfrequencies
$\Omega_q=\sqrt{\omega_P^2+\omega_q^2}$, where $\omega_P=\sqrt{
4\pi{n}_0e^2/m}$ is the fermion plasma frequency. The vector field
$\vec{A}(\vec{q},\lambda)$ can be written in terms of
these new eigenmodes as, 
\begin{equation}
\label{aphoton}
\vec{A}(\vec{q},\lambda)=\left({2\pi\over\Omega_q}\right)^{1\over2}
\vec{\xi}(\vec{q},\lambda)(b_{\lambda}(\vec{q})+b^+_{\lambda}(-\vec{q})).
\end{equation}
\end{mathletters}

  Integrating out the fermion fields $f(f^+)$'s we obtain
an action in terms of $\rho_{\vec{k}\sigma}(\vec{q})$ and
$\lambda_{\vec{k}\sigma}(\vec{q})$ fields,
\begin{mathletters}
\label{action1}
\begin{eqnarray}
\label{lan2}
S & = & {1\over\beta}F_0-Trln\left[\hat{1}-\hat{G}_0\hat{\lambda}\right]
-\sum_{\vec{k},\vec{q},i\omega_n,\sigma}i\lambda_{\vec{k}\sigma}
(\vec{q},-i\omega_n)\rho_{\vec{k}\sigma}(\vec{q},i\omega_n)+{1\over{L}^d}
\sum_{\vec{q},i\omega_n,\vec{k},\vec{k}'}v(q)\rho_{\vec{k},i\omega_n}
(\vec{q})\rho_{\vec{k}'}(-\vec{q},-i\omega_n)
\\ \nonumber
 &  &  -{\sqrt{2}e\over{L}^{d/2}}\sum_{\vec{k},\vec{q}\lambda,i\omega_n}
({\vec{k}.\vec{A}(\vec{q},\lambda,i\omega_n)\over{m}})
\rho_{\vec{k}}(\vec{q},-i\omega_n)+S'_{photon},
\end{eqnarray}
where $F_0$ is the free energy for an non-interacting Fermi gas.
$\hat{G}_0$ and $\hat{\lambda}$ are infinite matrices in
wave vector and frequency space, with matrix elements given by
\begin{equation}
\label{green}
\left[\hat{G}_0\right]_{k\sigma,k'\sigma'}=\delta_{\sigma\sigma'}
\delta_{k,k'}g_0(k)\; \;\;\;
g_0(k)={1\over{i}\omega_n-\xi_{\vec{k}}},
\end{equation}
and
\begin{equation}
\label{lambda}
\left[\hat{\lambda}\right]_{k\sigma,k'\sigma'}=\delta_{\sigma\sigma'}
{i\over\sqrt{\beta}}\lambda_{{\vec{k}+\vec{k}'\over2}\sigma}
(\vec{k}-\vec{k}',i\omega_n-i\omega_{n'})=\delta_{\sigma\sigma'}
{i\over\sqrt{\beta}}\lambda_{{k+k'\over2}\sigma}(k-k'),
\end{equation}
\end{mathletters}
where $k=(\vec{k},i\omega_n)$ and $\xi_{\vec{k}}=k^2/2m-\mu$. The
$Trln\left[1-G_0\lambda\right]$ term can be expanded in
a power series of $i\lambda_{\vec{k}\sigma}(q)$ field,
\[
Trln\left[\hat{1}-\hat{G}_o\hat{\lambda}\right]=-Tr\left[\hat{G}_0
\hat{\lambda}\right]-{1\over2}Tr\left[\hat{G}_0\hat{\lambda}\right]^2
-{1\over3}Tr\left[\hat{G}_0\hat{\lambda}\right]^3+O(\hat{\lambda}^4).  \]
Keeping terms to second order in $\hat{\lambda}$ (Gaussian approximation), 
we obtain
\begin{equation}
\label{gauss}
Trln\left[\hat{1}-\hat{G}_0\hat{\lambda}\right]\sim{1\over2
\beta}\sum_{k,q,\sigma}g_0(k+{q\over2})g_0(k-{q\over2})
\lambda_{k\sigma}(q)\lambda_{k\sigma}(-q).
\end{equation}
Notice that the first order term in $\hat{\lambda}$ gives the usual
Hartree self-energy and is excluded in our Hamiltonian. The
$i\lambda_{\vec{k}\sigma}(q)$ fields in Action \ (\ref{lan2}) can
be integrated out in Gaussian approximation, resulting in an quadratic
action $S$ in terms of $\rho_{\vec{k}}(q)$, $\sigma_{\vec{k}}(q)$ and 
photon fields only. We obtain $S=S_{\rho-p}+S_{\sigma}$, where
\begin{mathletters}
\begin{eqnarray}
\label{srho}
S_{\rho-p} & = & {1\over2}\sum_{\vec{k},\vec{k}',\vec{q},i\omega_n}
\left[-{1\over\chi_{0\vec{k}}(\vec{q},i\omega_n)}(\delta_{\vec{k},
\vec{k}'})+{2v(q)\over{L}^d}\right]\rho_{\vec{k}}(\vec{q},i\omega_n)
\rho_{\vec{k}'}(-\vec{q},-i\omega_n)  \\  \nonumber
& &  -{\sqrt{2}e\over{L}^{d/2}}\sum_{\vec{k},\vec{q}\lambda,i\omega_n}
({\vec{k}.\vec{A}(\vec{q},\lambda,i\omega_n)\over{m}})
\rho_{\vec{k}}(\vec{q},-i\omega_n)+S'_{photon}
\end{eqnarray}
and
\begin{equation}
\label{ssigma}
S_{\sigma}=-{1\over2}\sum_{\vec{k},\vec{q},i\omega_n}
{1\over\chi_{0\vec{k}}(\vec{q},i\omega_n)}\sigma_{\vec{k}}(\vec{q},i\omega_n)
\sigma_{\vec{k}}(-\vec{q},-i\omega_n),
\end{equation}
where
\begin{equation}
\label{x0}
\chi_{0\vec{k}}(\vec{q},i\omega_n)={1\over\beta}\sum_{i\Omega_n}
g_0(\vec{k}+\vec{q}/2,i\omega_n+i\Omega_n)g_0(\vec{k}-\vec{q}/2,i\Omega_n)
={n_{\vec{k}-\vec{q}/2}-n_{\vec{k}+\vec{q}/2}\over
i\omega_n-{\vec{k}.\vec{q}\over{m}}},
\end{equation}
\end{mathletters}
$n_{\vec{k}}=\theta(-\xi_{\vec{k}})$ at zero temperature is the free
fermion occupation number. $S_{\rho-p}$ and $S_{\sigma}$ can be 
expressed in terms of canonical boson fields by introducing
\begin{mathletters}
\label{can}
\begin{eqnarray}
\label{can1}
\rho_{\vec{k}}(\vec{q},i\omega_n) & = & \sqrt{|\Delta_{\vec{k}}(\vec{q})|}
\left(\theta(\Delta_{\vec{k}}(\vec{q}))b^+_{\vec{k}}(\vec{q},i\omega_n)
+\theta(-\Delta_{\vec{k}}(\vec{q}))b_{\vec{k}}(-\vec{q},-i\omega_n)\right),
\\  \nonumber
\sigma_{\vec{k}}(\vec{q},i\omega_n) & = & \sqrt{|\Delta_{\vec{k}}(\vec{q})|}
\left(\theta(\Delta_{\vec{k}}(\vec{q}))s^+_{\vec{k}}(\vec{q},i\omega_n)
+\theta(-\Delta_{\vec{k}}(\vec{q}))s_{\vec{k}}(-\vec{q},-i\omega_n)\right),
\end{eqnarray}
where $\Delta_{\vec{k}}(\vec{q})=n_{\vec{k}-\vec{q}/2}-n_{\vec{k}+
\vec{q}/2}$. Correspondingly, we also have
\begin{eqnarray}
\label{can2}
\rho_{\vec{k}}(-\vec{q},-i\omega_n) & = & \sqrt{|\Delta_{\vec{k}}(\vec{q})|}
\left(\theta(\Delta_{\vec{k}}(\vec{q}))b_{\vec{k}}(\vec{q},i\omega_n)
+\theta(-\Delta_{\vec{k}}(\vec{q}))b^+_{\vec{k}}(-\vec{q},-i\omega_n)\right),
\\  \nonumber
\sigma_{\vec{k}}(-\vec{q},-i\omega_n) & = & \sqrt{|\Delta_{\vec{k}}(\vec{q})|}
\left(\theta(\Delta_{\vec{k}}(\vec{q}))s_{\vec{k}}(\vec{q},i\omega_n)
+\theta(-\Delta_{\vec{k}}(\vec{q}))s^+_{\vec{k}}(-\vec{q},-i\omega_n)\right).
\end{eqnarray}
\end{mathletters}
Putting eqs. \ (\ref{can}) back into $S_{\rho-p}$, we obtain after 
some straightforward manipulations,
\begin{mathletters}
\label{actions}
\begin{eqnarray}
\label{sa}
S_{\rho-p} & = & {1\over2}\sum_{\vec{k},\vec{q},i\omega_n}(-i\omega_n+
{|\vec{k}.\vec{q}|\over{m}})b^+_{\vec{k}}(\vec{q},i\omega_n)
b_{\vec{k}}(\vec{q},i\omega_n)+{1\over2}\sum_{\vec{q}\lambda,i\omega_n}
(-i\omega_n+\Omega_q)b^+_{\lambda}(\vec{q},i\omega_n)
b_{\lambda}(\vec{q},i\omega_n) \\  \nonumber
& & +{1\over{L}^d}\sum_{\vec{k},\vec{k}',\vec{q},i\omega_n}v(q)
\sqrt{|\Delta_{\vec{k}}(\vec{q})\Delta_{\vec{k}'}(\vec{q})|}
\theta(\Delta_{\vec{k}}(\vec{q}))\theta(\Delta_{\vec{k}'}(\vec{q}))
\times\left(b^+_{\vec{k}}(\vec{q},i\omega_n)b_{\vec{k}'}(\vec{q},
i\omega_n)\right.   \\  \nonumber
& & \left.+b^+_{\vec{k}}(\vec{q},i\omega_n)b^+_{-\vec{k}'}(-\vec{q},
-i\omega_n)+b_{-\vec{k}}(-\vec{q},-i\omega_n)b_{\vec{k}'}(\vec{q},
i\omega_n)+b_{-\vec{k}}(-\vec{q},-i\omega_n)b^+_{-\vec{k}'}
(-\vec{q},-i\omega_n)\right)   \\  \nonumber
& & -{1\over{L}^{d/2}}\sum_{\vec{k},\vec{q}\lambda,i\omega_n}m_{\vec{k}}
(\vec{q},\lambda)\sqrt{|\Delta_{\vec{k}}(\vec{q})|}\theta(\Delta_{\vec{k}}
(\vec{q}))(b^+_{\vec{k}}(\vec{q},i\omega_n)-b_{-\vec{k}}(-\vec{q},-i\omega_n))
\\  \nonumber
& & \times(b_{\lambda}(\vec{q},i\omega_n)+b^+_{\lambda}(-\vec{q},-i\omega_n)),
\end{eqnarray}
and
\begin{equation}
\label{sb}
S_{\sigma}={1\over2}\sum_{\vec{k},\vec{q},i\omega_n}(-i\omega_n+
{|\vec{k}.\vec{q}|\over{m}})s^+_{\vec{k}}(\vec{q},i\omega_n)
s_{\vec{k}}(\vec{q},i\omega_n),
\end{equation}
\end{mathletters}
where
\[
m_{\vec{k}}(\vec{q},\lambda)={2e\over{m}}({\pi\over\Omega_q})^{1\over2}
\vec{k}.\vec{\xi}(\vec{q},\lambda).  \]
$S_{\rho-p}$ describes a coupled system of (fermion) {\em density} 
particle-hole pair excitations and photons, whereas $S_{\sigma}$ describes 
a system of free (fermion) {\em spin} particle-hole excitations. 
The density- and 
spin- particle-hole pair excitations are described by boson fields
$b(b^+)_{\vec{k}}(\vec{q})$ and $s(s^+)_{\vec{k}}(\vec{q})$, respectively,
satisfying usual boson commutation rules 
$[b(s)_{\vec{k}}(\vec{q}),b(s)^+_{\vec{k}'}(\vec{q}')]=
\delta_{\vec{k}\vec{k}'}\delta_{\vec{q}\vec{q}'}$ and
$[b(s)_{\alpha},b(s)_{\beta}]=[b(s)^+_{\alpha},b(s)^+_{\beta}]=0$, etc. 
In this form the dynamics of the original fermion system is described 
completely in terms of charge- and spin- boson fields(bosonized). 

   Notice that we have restricted ourselves to the Gaussian 
approximation in deriving $S_{\rho-p}$ and $S_{\sigma}$. Higher order
interaction terms between bosons will appear in a cumulant expansion of 
the $\lambda_{\vec{k}}(q)$ fields\cite{b4}. The convergence of the 
cumulant expansion is formally controlled in our model by the small 
parameter $\epsilon\sim(\Lambda/k_F)$, where a term of order 
$[b^mb^{+(n-m)}]$ in the cumulant expansion has to leading order a term 
$\sim\epsilon^{2(d-1)\times(n-1)}$ in each increasing order of cumulant expansion\cite{cc,ng1}. Notice however that the smallness of 
$\epsilon$ does not guarantee that all calculated physical quantities 
will converge uniformly in the cumulant expansion. In particular, we 
shall see that infra-red divergence appears in higher-order 
expansion of the transverse gauge field\cite{gauge,khm}. 
We shall discuss in more detail the consequences of the infra-red
divergences in section V. Notice also that in the $\vec{q}\rightarrow0$ 
limit, $\Delta_{\vec{k}}(\vec{q})\rightarrow-\delta(\epsilon_{\vec{k}}-\mu)
({\vec{k}.\vec{q}\over{m}})$ and the usual "tomographic" bosonization 
procedure based on subdivision of Fermi surface into disjoint patches 
at small $\vec{q}$ is recovered\cite{b1,b2,b3}. Our bosonization
procedure can be viewed as a generalization of the tomographic
bosonization method for small wave vector $\vec{q}$ to arbitrary 
values of $|\vec{q}|<\Lambda$.

  It is straightforward to show that the Gaussian approximation is 
essentially the same as usual RPA theory for interacting
fermions\cite{ng1,b1,b2,b3}, except the additional assumption that 
particle-hole excitations can be treated as independent bosons in 
bosonization theory\cite{ng1}. Note that particle-hole pairs 
are not all {\em independent} in a fermion system because of the Pauli 
exclusion principle.  The assumption of independent bosons is a major
approximation in the Gaussian theory in two- and three- dimensions. 
This is in contrast to what happens in one dimension where the 
entire particle-hole excitation spectrum can be represented 
rigorously by bosons\cite{b1} when the fermion spectrum is linearized 
near the Fermi surface. 

   Despite the approximations in the Gaussian theory, the bosonized 
form of the action has the advantage that within the approximation the 
full excitation spectrum as well as the wavefunctions 
of the system can be obtained easily. This allows us to study the 
properties of the system in great detail, as we shall see
in the following.

\section{density particle-hole excitation spectrum}
   The eigenstates and eigenvalue spectrum described by
the action $S_{\rho-p}$ can be obtained by diagonalizing
the bosonized action \ (\ref{sa}) using a generalized
Bogoliubov transformation. We introduce for each wave vector
$\vec{q}$ the Bogoliubov transformation\cite{b5}
\begin{mathletters}
\begin{eqnarray}
\label{bg}
b_{k}(\vec{q}) & = & \sum_{k'}\left[\alpha_{kk'}^>\gamma_{k'}(\vec{q})+
\beta_{kk'}^>\gamma^+_{-k'}(-\vec{q})\right],  \\  \nonumber
b_{-k}(-\vec{q}) & = & \sum_{k'}\left[\alpha_{kk'}^<
\gamma_{-k'}(-\vec{q})+\beta_{kk'}^<\gamma^+_{k'}
(\vec{q})\right],
\end{eqnarray}
where $k=\vec{k},\lambda(k'=\vec{k}',\lambda')$, and with 
correspondingly,
\begin{eqnarray}
\label{bgi}
\gamma_{k}(\vec{q}) & = & \sum_{k'}\left[\alpha^{>*}_{k'k}b_{k'}(\vec{q})-\beta_{k'k}^<b^+_{-k'}
(-\vec{q})\right],  \\  \nonumber
\gamma_{-k}(-\vec{q}) & = & \sum_{k'}\left[\alpha^{<*}_{k'k}b_{-k'}(-\vec{q})-\beta_{k'k}^>b^+_{k'}
(\vec{q})\right],
\end{eqnarray}
\end{mathletters}
where we require that the $\gamma(\gamma^+)_{k}(\vec{q})$ operators
diagonize the Hamiltonian, i.e.
\begin{equation}
\label{hbo}
H_{\rho-p}=\sum_{k}E_{k}(\vec{q})\gamma^+_{k}(\vec{q})\gamma_{k}(\vec{q})
+E_G,   
\end{equation}
where $E_{k}(\vec{q})$ are the eigen-energies and $E_{G}$
is the ground-state energy of the system. The eigenstates may represent
dressed particle-hole excitations ($k=\vec{k}$) or dressed photons
($k=\lambda$). Notice that additional collective modes may appear in the
system and is also included in the sum $\sum_{k}$. The matrix
elements $\alpha$ and $\beta$ satisfies the orthonormality condition
\begin{eqnarray}
\label{on}
\sum_{k"}\left[\alpha_{kk"}^{>(<)}\alpha^{>(<)*}_{k'k"}-
\beta_{kk"}^{>(<)}\beta^{>(<)*}_{k'k"}\right] 
& = & \delta_{kk'},   \\  \nonumber
\sum_{k"}\left[\alpha_{kk"}^<\beta_{k'k"}^>-\beta_{kk"}^<\alpha_{k'k"}^>\right]
 & = & 0.
\end{eqnarray}
   Writing down the equation of motions for $b_{k}(\vec{q})$'s
in terms of $\gamma(\gamma^+)_{k}(\vec{q})$'s\cite{b5}, we obtain
the Bogoliubov equations
\begin{mathletters}
\label{bogo}
\begin{eqnarray}
\label{b1}
(E_{k'}(\vec{q})-{|\vec{k}.\vec{q}|\over{m}})\alpha_{\vec{k}k'}^{>(<)}
& = & -\theta(\Delta_{\vec{k}}(\vec{q})\sqrt{|\Delta_{\vec{k}}(\vec{q})|}
\left[{+(-)1\over{L}^{d/2}}\sum_{\lambda}m_{\vec{k}}(\vec{q},\lambda)
(\alpha^{>(<)}_{\lambda{k}'}+\beta^{<(>)*}_{\lambda{k}'})\right. \\  \nonumber
&  &  -\left.{2v(q)\over{L}^d}
\sum_{\vec{k}"}\theta(\Delta_{\vec{k}"}(\vec{q}))\sqrt{|\Delta_{\vec{k}"}
(\vec{q})|}(\alpha_{\vec{k}"k'}^{>(<)}+\beta^{<(>)*}_{\vec{k}"k'})\right],
\end{eqnarray}
\begin{eqnarray}
\label{b2}
(E_{k'}(\vec{q})+{|\vec{k}.\vec{q}|\over{m}})\beta_{\vec{k}k'}^{<(>)}
& = & -\theta(\Delta_{\vec{k}}(\vec{q})\sqrt{|\Delta_{\vec{k}}(\vec{q})|}
\left[{+(-)1\over{L}^{d/2}}\sum_{\lambda}m_{\vec{k}}(\vec{q},\lambda)
(\alpha^{>(<)*}_{\lambda{k}'}+\beta^{<(>)}_{\lambda{k}'})\right. \\  \nonumber
&  & \left.+{2v(q)\over{L}^d}
\sum_{\vec{k}"}\theta(\Delta_{\vec{k}"}(\vec{q}))\sqrt{|\Delta_{\vec{k}"}
(\vec{q})|}(\alpha^{>(<)*}_{\vec{k}"k'}+\beta^{<(>)}_{\vec{k}"k'})\right],
\end{eqnarray}
and
\begin{equation}
\label{b3}
(E_{k'}(\vec{q})-\Omega_q)\alpha^{>(<)}_{\lambda{k}'}=
{-(+)1\over{L}^{d/2}}\sum_{\vec{k}"}m_{\vec{k}"}(\vec{q},\lambda)
\theta(\Delta_{\vec{k}"}(\vec{q}))\sqrt{|\Delta_{\vec{k}"}
(\vec{q})|}(\alpha^{>(<)}_{\vec{k}"k'}-\beta^{<(>)*}_{\vec{k}"k'}),
\end{equation}
\begin{equation}
\label{b4}
(E_{k'}(\vec{q})+\Omega_q)\beta^{<(>)}_{\lambda{k}'}={+(-)1\over{L}^{d/2}}
\sum_{\vec{k}"}m_{\vec{k}"}(\vec{q},\lambda)\theta(\Delta_{\vec{k}"}
(\vec{q}))\sqrt{|\Delta_{\vec{k}"}(\vec{q})|}(\alpha^{>(<)*}_{\vec{k}"k'}
-\beta^{<(>)}_{\vec{k}"k'}).
\end{equation}
\end{mathletters}
   
In general we find that there exists two kinds of solutions to these
equations: (i)particle-hole continuum, with $E_{\vec{k}}(\vec{q})=|
\vec{k}.\vec{q}|/m$, and (ii)collective modes, including (renormalized)
photons with energy $E_{\lambda}(\vec{q})$ determined by the eigenvalue
equation $E_{\lambda}(\vec{q})^2-\Omega_q^2-{4\pi{e}^2\over{m}^2}
\chi_t(\vec{q},E_{\lambda}(\vec{q}))=0$, and plasmons, with energy
$E_0(\vec{q})$ given by the eigenvalue equation $1-v(q)
\chi_0(\vec{q},E_0(\vec{q}))=0$, where 
\[
\chi_t(\vec{q},\omega)={2\over{L}^d(d-1)}\sum_{\vec{k}}
(\vec{k}_t.\vec{k}_t)\chi_{0\vec{k}}(\vec{q},\omega),  \]
is the paramagnetic transverse current susceptibility, 
$\vec{k}_t=\vec{k}-\hat{q}(\vec{k}.\hat{q})$ and 
\[
\chi_0(\vec{q},\omega)={2\over{L}^d}\sum_{\vec{k}}
\chi_{0\vec{k}}(\vec{q},\omega).  \]
is the Lindhard function. The energies 
of these collective modes are outside the particle-hole continuum. We
now consider the solutions in more detail. First we consider the 
particle-hole excitations. After some lengthy algebras, we obtain,
\begin{mathletters}
\label{sol}
\begin{eqnarray}
\label{sol1}
\alpha^>_{\vec{k}\vec{k}'} & = & \delta_{\vec{k}\vec{k}'}+P{\theta(\Delta
_{\vec{k}}(\vec{q}))\theta(\Delta_{\vec{k}'}(\vec{q}))\sqrt{|
\Delta_{\vec{k}}(\vec{q})\Delta_{\vec{k}'}(\vec{q})|}\over{L}^d
({|\vec{k}'.\vec{q}|\over{m}}-{|\vec{k}.\vec{q}|\over{m}})}\times\left(
2v_{eff}(q,|\vec{k}'.\vec{q}|/m)-\vec{k}_t.\vec{A}_{eff}(\vec{q},\vec{k}')
\right),   \\  \nonumber
\beta^<_{\vec{k}\vec{k}'} & = & {-\theta(\Delta_{\vec{k}}(\vec{q}))
\theta(\Delta_{\vec{k}'}(\vec{q}))\sqrt{|\Delta_{\vec{k}}(\vec{q})
\Delta_{\vec{k}'}(\vec{q})|}\over{L}^d({|\vec{k}'.\vec{q}|\over{m}}+
{|\vec{k}.\vec{q}|\over{m}})}\times\left(2v_{eff}(q,-|\vec{k}'.\vec{q}|/m)
+\vec{k}_t.\vec{A}_{eff}^*(\vec{q},\vec{k}')\right),  
 \\  \nonumber
\alpha^>_{\lambda\vec{k}'} & = & -{(|\vec{k}'.\vec{q}|/m+\Omega_q){
2e\over{m}}({\pi\over\Omega_q})^{1/2}\theta(\Delta_{\vec{k}'}(\vec{q}))
\sqrt{|\Delta_{\vec{k}'}(\vec{q})|}\over{L}^{d/2}((|\vec{k}'.\vec{q}|/m)^2-
\Omega_q^2-{4\pi{e}^2\over{m}^2}\chi_t(\vec{q},|\vec{k}'.\vec{q}|/m)}
(\vec{\xi}_{\lambda}(\vec{q}).\vec{k}_t'),  \\  \nonumber
\beta^<_{\lambda\vec{k}'} & = & {(|\vec{k}'.\vec{q}|/m-\Omega_q){
2e\over{m}}({\pi\over\Omega_q})^{1/2}\theta(\Delta_{\vec{k}'}(\vec{q}))
\sqrt{|\Delta_{\vec{k}'}(\vec{q})|}\over{L}^{d/2}
((|\vec{k}'.\vec{q}|/m)^2-\Omega_q^2-{4\pi{e}^2\over{m}^2}\chi_t
(\vec{q},-|\vec{k}'.\vec{q}|/m)}(\vec{\xi}_{\lambda}(\vec{q}).\vec{k}_t'),
\end{eqnarray}
for the particle-hole continuum spectrum $\vec{k}'$, where
$v_{eff}(q,\omega)=v(q)/(1-v(q)\chi_0(\vec{q},\omega)$ is the RPA effective
interaction, and $\vec{A}_{eff}(\vec{q},\vec{k}')=({8\pi{e}^2\over{m}^2}
\vec{k}'_t)/((|\vec{k}'.\vec{q}|/m)^2-\Omega_q^2-{4\pi{e}^2\over{m}^2}
\chi_t(\vec{q},|\vec{k}'.\vec{q}|/m))$. We also obtain $\alpha(\beta)^<_{\lambda\vec{k}'}=-\alpha(\beta)^>_{\lambda\vec{k}'}$, 
and $\alpha(\beta)^<_{\vec{k}
\vec{k}'}=\alpha(\beta)^>_{\vec{k}\vec{k}'}$. 

   For the collective modes, we obtain for the longitudinal (plasmon)
mode,
\begin{eqnarray}
\label{sol2}
\alpha_{\vec{k}0}^> & = & 
{1\over{L}^{d/2}}{2\theta(\Delta_{\vec{k}}(\vec{q}))
\sqrt{|\Delta_{\vec{k}}(\vec{q})|}\over(E_0(\vec{q})-{|\vec{k}.\vec{q}|
\over{m}})\left[-{\partial\chi_0(q,\omega)\over\partial\omega}\right]
_{\omega=E_0(\vec{q})}^{1\over2}},  \\  \nonumber
\beta_{\vec{k}0}^{<*} & = & 
-{1\over{L}^{d/2}}{2\theta(\Delta_{\vec{k}}
(\vec{q}))\sqrt{|\Delta_{\vec{k}}(\vec{q})|}\over(E_0(\vec{q})+
{|\vec{k}.\vec{q}|\over{m}})\left[-{\partial\chi_0(q,\omega)
\over\partial\omega}\right]_{\omega=E_0(\vec{q})}^{1\over2}}, \\  \nonumber
\alpha_{\lambda0}^> & = & \beta_{\lambda0}^{<*} = 0,
\end{eqnarray}
and $\alpha(\beta)^<_{\vec{k}0}=\alpha(\beta)^>_{\vec{k}0}$,
$\alpha(\beta)^<_{\lambda0}=0$. Notice that in bosonization theory, 
plasmon arises from non-perturbative interaction effect and cannot be 
obtained from analytic continuation of the non-interacting boson modes.
Similarly, we also obtain for the (renormalized) photon modes,
\begin{eqnarray}
\label{sol3}
\alpha^>_{\vec{k}\lambda} & = & -{\theta(\Delta_{\vec{k}}(\vec{q}))
\sqrt{|\Delta_{\vec{k}}(\vec{q})|}\over{E}_{\lambda}(\vec{q})-
{|\vec{k}.\vec{q}|\over{m}}}\left({\vec{k}.\xi_{\lambda}(\vec{q})C(q)
\over{L}^{d/2}}\right),  \\  \nonumber
\beta^{<*}_{\vec{k}\lambda} & = & -{\theta(\Delta_{\vec{k}}(\vec{q}))
\sqrt{|\Delta_{\vec{k}}(\vec{q})|}\over{E}_{\lambda}(\vec{q})+
{|\vec{k}.\vec{q}|\over{m}}}\left({\vec{k}.\xi_{\lambda}(\vec{q})C(q)
\over{L}^{d/2}}\right),  \\  \nonumber
\alpha^>_{\lambda\lambda'} & = & {\delta_{\lambda\lambda'}\over
(E_{\lambda}(\vec{q})-\Omega_q)}{e\over{m}}
\left({\pi\over\Omega_q}\right)^{1\over2}\chi_t(\vec{q},E_{\lambda}
(\vec{q}))C(q),  \\  \nonumber
\beta^{<*}_{\lambda\lambda'} & = & -{\delta_{\lambda\lambda'}\over
(E_{\lambda}(\vec{q})+\Omega_q)}{e\over{m}}
\left({\pi\over\Omega_q}\right)^{1\over2}\chi_t(\vec{q},E_{\lambda}
(\vec{q}))C(q), 
\end{eqnarray}
where
\[
C(q)=\left({({4\pi{e}^2\over{m}^2})\over2E_{\lambda}(\vec{q})-
({4\pi{e}^2\over{m}^2})[{\partial\chi_t(\vec{q},\omega)\over
\partial\omega}]_{\omega=E_{\lambda}(\vec{q})}}\right)^{1\over2},
\]
\end{mathletters}
and with $\alpha(\beta)_{\lambda'\lambda}^<=\alpha
(\beta)_{\lambda'\lambda}^>$ and $\alpha(\beta)_{\vec{k}\lambda}^<=
-\alpha(\beta)_{\vec{k}\lambda}^>$. 

   Next we examine the solutions of the bosonized Hamiltonian in the
$e^2\rightarrow\infty$ limit. First we consider the collective modes.
Using the result that $\chi_0(\vec{q},\omega)\rightarrow{n_0q^2\over{m}
\omega^2}$, and $\chi_t(\vec{q},\omega)\rightarrow{2n_0q^2\over{d}\omega^2}
<\epsilon>$ in the limit $\omega>>k_Fq/m$, where $n_0$ is the fermion 
density\cite{mahan} and $<\epsilon>$ is the average kinetic energy 
per fermions in the free fermion ground state, it is easy to see 
that in the limit $e^2\rightarrow\infty$, the collective mode
frequencies $E_0(q),E_{\lambda}(q)$ all approaches the plasma
frequency $\omega_P=({4\pi{n}_0e^2\over{m}})^{1\over2}$. Notice that
$\omega_P\rightarrow\infty$ as $e^2\rightarrow\infty$, indicating that 
the collective modes are outside the physical spectrum in this limit. 

  Despite the vanishing of collective excitations in the physical
spectrum, the particle-hole excitation spectrum with excitation energies
$|\vec{k}.\vec{q}|/m$ survives in bosonization theory in
the limit $e^2\rightarrow\infty$. In this limit
$v_{eff}(q,|\vec{k}'.\vec{q}|/m)\rightarrow-1/\chi_0(q,|\vec{k}'.
\vec{q}|/m)$ and $\vec{k}.\vec{A}_{eff}(\vec{q},\vec{k}')\rightarrow
-2\vec{k}_t.\vec{k}_t'/(mn_0+\chi_t(q,|\vec{k}'
.\vec{q}|/m))$. The coefficients $\alpha_{\vec{k}\vec{k}'}$'s and $\beta_{\vec{k}\vec{k}'}$'s remain regular, indicating
that the particle-hole excitation spectrum survives under effect of 
confinement. It is instructive to examine the charge and (transverse) 
current fluctuations carried by the (eigen)-particle-hole excitations by 
examining the commutators
$[\rho(\vec{q}),\gamma_{\vec{k}}(\vec{q}')]$ and $[\vec{j}_t
(\vec{q}),\gamma_{\vec{k}}(\vec{q}')]$, where
\[
\vec{j}_t(\vec{q})=\sum_{\vec{k}\sigma}{e\vec{k}_t\over{m}}
\rho_{\vec{k}\sigma}(\vec{q})-(L^{d/2})\sum_{\lambda}{e^2n_0\over{m}}\vec{A}
(-\vec{q},\lambda), \]
is the transverse current operator. In particular, we expect 
that these commutators should vanish in the $e^2\rightarrow\infty$ limit, 
where charge and current fluctuations are forbidden. 
Using Eqs.\ (\ref{can}), \ (\ref{bgi}), \ (\ref{sol1}) and usual boson
commutation rules, it is straightforward to show that
\begin{equation}
\label{cchar}
[e\rho(\vec{q}),\gamma_{\vec{k}}(\vec{q}')]=\delta_{\vec{q}\vec{q}'}
\theta(\Delta_{\vec{k}}
(\vec{q}))\sqrt{|\Delta_{\vec{k}}(\vec{q})|}\times{-e\over1-v(q)
\chi_0(\vec{q},|\vec{k}.\vec{q}|/m)},
\end{equation} 
and
\begin{equation}
\label{ccurr}
[\vec{j}_t(\vec{q}),\gamma_{\vec{k}}(\vec{q}')]=\delta_{\vec{q}\vec{q}'}
{-\sqrt{2}e\over{m}}\theta(\Delta_{\vec{k}}
(\vec{q}))\sqrt{|\Delta_{\vec{k}}(\vec{q})|}\times\left({
{|\vec{k}.\vec{q}|\over{m}}-(cq)^2\over
{|\vec{k}.\vec{q}|\over{m}}-\Omega_q^2-{4\pi{e}^2\over{m}^2}\chi_t
(\vec{q},{|\vec{k}.\vec{q}\over{m}})}\right),
\end{equation}
   both vanishes in the limit $e^2\rightarrow\infty$.

   Let us summarize our results obtained so far from bosonization 
theory. Within the Gaussian approximation, we obtain a RPA-like 
charge excitation spectrum with both collective modes
and particle-hole excitations, and a free-fermion type spin particle-hole
excitation spectrum. As the coupling constant $e^2$ increases, the 
energies of the charge collective modes rise continuously to infinity 
whereas both the charge and spin particle-hole excitation spectrum 
remains intact. The charge and current fluctuations carried by the 
charge particle-hole excitations are projected out gradually as $e^2$ 
increases, resulting in {\em chargeless} particle-hole excitations 
in the confinement limit $e^2\rightarrow\infty$. 
Notice that within the Gaussian approximation, the confinement state 
analytically continues to the usual Fermi liquid state and there is no 
phase transition in between. The theory thus suggests that the 
confinement state of a gas of fermions is a fermion liquid state
formed by spin $S=1/2$ and {\em chargeless}-quasi-particles. It also 
suggests that this is a rather unusual fermion liquid state, since 
bare fermions in the system carries spin $1/2$ and charge $e$, and the
quasi-particles must have vanishing overlap with bare fermions if they 
carry zero charge. The nature of the liquid of chargeless fermions will 
be examined in more details in next section, where we shall construct 
explicitly "chargeless" fermion operators that describe the dynamics of 
the system in the limit $e^2\rightarrow \infty$. The limitations of 
Gaussian theory will be uncovered in the process. 

\section{single-particle properties and low-energy effective Lagrangian}
    In bosonization theory for one-dimensional systems, the single 
particle properties of the system can be determined once a rigorous
representation of the single-particle operator in terms of density 
operators $\rho_L(q)$ and $\rho_R(q)$ are obtained\cite{b1}. In higher
dimensions, this procedure becomes inadequate for two reasons: (1)The
corresponding procedure requires that the bosonized representation of
single-particle operator $\psi_{b\sigma}(\vec{r})$ satisfies the commutation 
relations 
\[
[\psi_{b\sigma}(\vec{r}),\rho_{\vec{k}}(\vec{q})]=e^{-i(\vec{k}+\vec{q}/2).
\vec{r}}\int{d}^dr'e^{i(\vec{k}-\vec{q}/2).\vec{r}'}\psi_{b\sigma}(\vec{r}')
\]
for all possible momenta $\vec{k}$ and $\vec{q}$. We have not 
been able to find a representation which satisfies this
criteria\cite{b2,b3,b4,ng1}, and even if we can find such an 
representation, the theory would still be approximate because in 
dimensions higher than one, the boson representation using Wigner 
operators is not exact, (2)more importantly, unlike in one 
dimension where the elementary excitations are collective density- and
spin- wave modes, bosonization theory suggests that in dimensions 
higher than one the particle-hole excitation spectrum is fermi-liquid 
like, implying that fermionic quasi-particles exist in dimensions higher 
than one. It is important to construct directly quasi-particle 
operators in this case.

   To construct the quasi-particle operators we start from the equation 
of motion of the bare fermion operator $\psi_{\sigma}(\vec{r})=
{1\over{L}^{d/2}}\sum_{\vec{k}}e^{i\vec{k}.\vec{r}}f_{\vec{k}\sigma}$ at imaginary time,
\begin{eqnarray}
\label{emb}
{\partial\psi_{\sigma}(\vec{r})\over\partial\tau} 
& = & [H,\psi_{\sigma}(\vec{r})]  \\ \nonumber
& = & {1\over2m}\nabla^2\psi_{\sigma}
(\vec{r})-{1\over{L}^d}\sum_{\vec{q}}v(q)\rho(\vec{q})e^{-i\vec{q}.\vec{r}}
\psi_{\sigma}(\vec{r})-{ie\over{m}L^{d/2}}\sum_{\vec{q}\lambda}
e^{i\vec{q}.\vec{r}}\vec{A}(\vec{q},\lambda).\nabla\psi_{\sigma}
(\vec{r}).
\end{eqnarray}
   where we have replaced the fermion-density operator $\rho(\vec{r})$
by its expectation value $n_0$ in the diamagnetic term as we have done
in sections II and III in deriving \ (\ref{emb}). We shall discuss the
validity of this approximation at the end of this section.
In the bosonization approximation, the operators $\rho(\vec{q})$
and $\vec{A}(\vec{q},\lambda)$ can be decomposed as $\rho(\vec{q})=
\rho_{ph}(\vec{q})+\rho_{c}(\vec{q})$, and $\vec{A}(\vec{q},\lambda)
=\vec{A}_{ph}(\vec{q},\lambda)+\vec{A}_{c}(\vec{q},\lambda)$,  where
$v(q)\rho_{ph}(\vec{q})$ and ${e\vec{A}_{ph}(\vec{q},\lambda)
\over{c}}$ describes the coupling of the fermion to particle-hole 
excitations through the scalar and vector field fluctuations, 
respectively, whereas $v(q)\rho_c(\vec{q})$ and
${e\vec{A}_{c}(\vec{q},\lambda)\over{c}}$ describes coupling of the 
fermion to collective modes (plasmons and photons). It is
straightforward to obtain
\begin{mathletters}
\begin{eqnarray}
\label{rho1}
v(q)\rho_{ph}(q) & = & \sum_{\vec{k}}\sqrt{|\Delta_{\vec{k}}(\vec{q})|}
\theta(\Delta_{\vec{k}}(\vec{q}))\left[v_{eff}(q,{-|\vec{k}.\vec{q}|
\over{m}})\gamma^+_{\vec{k}}(\vec{q})+v_{eff}(q,{|\vec{k}.\vec{q}|
\over{m}})\gamma_{-\vec{k}}(-\vec{q})\right],   \\  \nonumber
v(q)\rho_c(\vec{q}) & = & L^{d/2}\left(-{\partial\chi_0(q,\omega)
\over\partial\omega}\right)^{-{1\over2}}_{\omega=E_0(\vec{q})}
\left[\gamma_0^+(\vec{q})+\gamma_0(-\vec{q})\right],
\end{eqnarray}
and
\begin{eqnarray}
\label{a1}
{e\vec{A}_{ph}(\vec{q},\lambda)\over{m}} & = & {-1\over\sqrt{2}L^{d/2}}
\vec{\xi}_{\lambda}(\vec{q})\sum_{k}\theta(\Delta_{\vec{k}}(\vec{q})
\sqrt{|\Delta_{\vec{k}}(\vec{q})|}\left[\vec{\xi}_{\lambda}(\vec{q}).
\vec{A}_{eff}(\vec{q},\vec{k})\gamma_{\vec{k}}(\vec{q})-
\vec{\xi}_{\lambda}(\vec{q}).\vec{A}_{eff}^*(\vec{q},\vec{k})
\gamma^+_{-\vec{k}}(-\vec{q})\right]  \\  \nonumber
{e\vec{A}_{c}(\vec{q},\lambda)\over{m}} & = & \sqrt{2}C(q)
\vec{\xi}_{\lambda}(\vec{q})\left(\gamma_{\lambda}(\vec{q})+
\gamma^+_{\lambda}(-\vec{q})\right).
\end{eqnarray}
\end{mathletters}
In particular, it is easy to see that in the $e^2\rightarrow\infty$ limit, 
the interaction between fermion and particle-hole excitations in both
longitudinal ($v(q)\rho_{ph}(q)$) and transverse
($e\vec{A}_{ph}(\vec{q},\lambda)/c)$ channels remains regular and finite, 
and divergences in the $e^2\rightarrow\infty$ limit appear only through 
the interactions between fermion and collective excitations. Notice that 
{\em infra-red} divergences in the interaction between fermion and 
particle-hole excitations also exist. However, they are not 
results of taking $e^2\rightarrow\infty$ and are not considered here. We 
shall discuss the effects of infra-red divergences in the last section.

  The divergence in interaction between fermions and collective modes
suggests that we have to eliminate these interactions first to study 
dynamics of (physical) fermion operators in the limit 
$e^2\rightarrow\infty$. To eliminate these interactions we look for 
a canonical transformation for the fermion operator\cite{mahan},
\begin{equation}
\label{qu1}
\psi_{P\sigma}(\vec{r})=e^{\phi(\vec{r})}\psi_{\sigma}(\vec{r}).
\end{equation}
with corresponding equation of motion,
\begin{equation}
\label{cano2}
[H,\psi_{P\sigma}(\vec{r})]=[H,e^{\phi(\vec{r})}]\psi_{\sigma}(\vec{r})
+e^{\phi(\vec{r})}[H,\psi_{\sigma}(\vec{r})],
\end{equation}
where we shall choose $\phi(\vec{r})$ such that $[H,e^{\phi}]$ cancels 
the interaction term between fermion and collective modes, i.e.
\[
[H,e^{\phi(\vec{r})}]\psi_{\sigma}(\vec{r})=e^{\phi(\vec{r})}\left[
{1\over{L}^d}\sum_{\vec{q}}v(q)\rho_c(\vec{q})e^{-i\vec{q}.\vec{r}}
\psi_{\sigma}(\vec{r})+{ie\over{m}L^{d/2}}\sum_{\vec{q}\lambda}
e^{i\vec{q}.\vec{r}}\vec{A}_c(\vec{q},\lambda).\nabla\psi_{\sigma}
(\vec{r})\right]. \]
Furthermore, we shall assume that $\phi(\vec{r})$ depends linearly
on the collective mode operators $\gamma_{0(\lambda)}(\vec{q})$ and
$\gamma_{0(\lambda)}^+(\vec{q})$'s, and that $H$ can be replaced by 
the bosonized Hamiltonian \ (\ref{hbo}) in evaluating the 
commutator $[H,e^{\phi}]$. Notice that the last assumption is 
justified in the $e^2\rightarrow\infty$ or $\vec{q}
\rightarrow0$ limit, where both the longitudinal and transverse 
collective mode excitations become exact.

 With these conditions, it is straightfoward to show that
$\phi(\vec{r})=\phi_l(\vec{r})+\vec{W}(\vec{r}).\nabla$, where 
\begin{eqnarray}
\label{qu2}
\phi_l(\vec{r}) & = & {1\over{L}^{d/2}}\sum_{\vec{q}}
{e^{-i\vec{q}.\vec{r}}\over{E}_0(\vec{q})
\left[-{\partial\chi_0(q,\omega)\over\partial\omega}\right]^{1\over2}}
(\gamma^+_0(\vec{q})-\gamma_0(-\vec{q})),  \\ \nonumber
\vec{W}(\vec{r}) & = & {\sqrt{2}i\over{L}^{d/2}}\sum_{\vec{q}\lambda}
{e^{i\vec{q}.\vec{r}}C(q)\over{E}_{\lambda}(\vec{q})}(\gamma_
{\lambda}^+(-\vec{q})-\gamma_{\lambda}(\vec{q}))
\vec{\xi}_{\lambda}(\vec{q}),
\end{eqnarray}
where $\phi_l$ and $\vec{W}$ describes the dressing of fermion by
(longitudinal) plasmon modes and (transverse) photon modes, respectively. 
Notice that $[\phi_l(\vec{r}),\vec{W}(\vec{r}').\nabla']=0$, because of
decoupling between transverse and longitudinal fluctuations.

To show that $\psi_{P\sigma}(\vec{r})$ describes chargeless fermions 
in the $e^2\rightarrow\infty$ limit, we examine the commutation
relations between $\psi_{P\sigma}(\vec{r})$ and density and 
(transverse) current operators. It is straightforward to show that
\begin{mathletters}
\label{comm}
\begin{equation}
\label{cles1}
[e\rho(\vec{q}),\psi_{P\sigma}(\vec{r})]=e\left({2\chi_0(q,E_0(\vec{q}))\over
E_0(\vec{q})\left[-{\partial\chi_0(q,\omega)\over\partial\omega}
\right]_{\omega=E_0(\vec{q})}}-1\right)e^{i\vec{q}.\vec{r}}
\psi_{P\sigma}(\vec{r}),
\end{equation}
and
\begin{equation}
\label{cles2}
[\vec{j}_t(\vec{q}),\psi_{P\sigma}(\vec{r})]={ie\over{m}}\left(1-
{2C(q)^2\over{E}_{\lambda}(\vec{q})}\left[\chi_t(\vec{q},E_{\lambda}(\vec{q}))+
n_0m\right]\right)e^{i\vec{q}.\vec{r}}\nabla_t\psi_{P\sigma}(\vec{r}).
\end{equation}
\end{mathletters}
where $\nabla_t=\nabla-\hat{q}(\hat{q}.\nabla)$. It is easy to see
that both commutators vanish in the limit $e^2\rightarrow\infty$, 
when $E_{0(\lambda)}(\vec{q})\rightarrow\omega_P\rightarrow\infty$. It
is also easy to see that the $\psi_{P\sigma}$ operator has the same 
commutation relation with spin density operator $\sigma(\vec{q})
=\sum_{\vec{k}}\sigma_{\vec{k}}(\vec{q})$ as bare fermion operator
$\psi_{\sigma}$. These results together imply that the dressed single 
particle operators $\psi_{P\sigma}(\vec{r})$'s represent spin $S=1/2$, 
'chargeless' fermions in the limit $e^2\rightarrow\infty$.

 To show that $\psi_{P\sigma}(\vec{r})$ and $\psi_{P\sigma}(\vec{r}')$ 
represent independent physical fermionic excitations in the system when
$\vec{r}\neq\vec{r}'$ we examine the commutation relation
between the dressed fermion operators. We obtain
\[
\{\psi_{P\sigma}(\vec{r}),\psi_{P\sigma'}^+(\vec{r}')\}
\sim{\sqrt{2}\over{n}_0(\pi|\vec{r}-\vec{r}'|)^{d-1}}
\hat{O}(\vec{r},\vec{r}'),   \]
in the limits $e^2\rightarrow\infty$ and $|\vec{r}-\vec{r}'|\rightarrow
\infty$, where 
\[
\hat{O}=({\vec{r}-\vec{r}'\over|\vec{r}-\vec{r}'|}).\left[
\psi^+_{\sigma'}(\vec{r}')e^{\phi(\vec{r})}(\nabla\psi_{\sigma}(\vec{r}))
e^{\phi^+(\vec{r}')}-(\nabla\psi^+_{\sigma'}(\vec{r}'))
e^{\phi(\vec{r})}\psi_{\sigma}(\vec{r})e^{\phi^+(\vec{r}')}\right].  \] 
The main contribution to the commutator 
comes from nonzero commutation relation between the plasmon cloud
around one chargeless fermion ($\phi_l(\vec{r})$) and the other bare 
fermion ($\psi_{\sigma'}(\vec{r}')$). The non-vanishing commutator between 
chargeless fermions separated by large distance reflects the nonlocal
nature of the chargeless fermion operator. Fortunately, the power law 
decay of the commutator between $\psi_{P\sigma}$ operators separated by
large distances at dimensions larger than one indicates that they can 
be used as starting point to construct independent single fermion 
operators when describing the dynamics of the system at long 
distances. Notice that in one dimension, such a construction is not 
possible because of distance independence of the commutation relation. 
In fact, the only fermionic operators that commute with density 
operator $\rho(\vec{q})$'s are the ladder operators\cite{b1} which raise 
or lower the number of particles in the system by one. There are only 
four independent ladder operators in the system, which change the number 
of left-going and right-going spin-$\sigma$ fermions and cannot be used 
to construct local quasi-particle excitations.

  The relation between the fermion liquid formed by the chargeless and 
bare fermions can be seen by examining the relation between the
two ground state expectation values 
$<\psi^+_{\sigma}(\vec{r})\psi_{\sigma}(\vec{r}')>$ and $<\psi^+_{P\sigma}(\vec{r})\psi_{P\sigma}(\vec{r}')>$. Assuming 
that the collective modes can be treated as independent excitations, we 
obtain to leading order in the projected Hilbert space,
\begin{mathletters}
\begin{equation}
\label{o1}
<\psi^+_{\sigma}(\vec{r})\psi_{\sigma}(\vec{r}')>\sim
{e}^{-2({\pi{m}e^2\over{n}_0})^{1\over2}
\int^{\Lambda}_{L^{-1}}d^dq{1\over{q}^2}(1-cos(\vec{q}.(\vec{r}-\vec{r}')))}
<\psi^+_{P\sigma}(\vec{r})\psi_{P\sigma}(\vec{r}')>.  
\end{equation}
The singular behaviour in the exponential factor comes from plasmon
contributions and photons do not contribute to this order. Notice
that the average density of bare fermions and dressed fermions
are the same and the fermi surface volume of the dressed fermions is 
exactly the same as that of the bare fermions. It is also easy to see
that
\begin{equation}
\label{o2}
<\psi^+_{P\sigma}(\vec{r})\psi_{\sigma}(\vec{r})>
\sim{e}^{-({\pi{m}e^2\over{n}_0})^{1\over2}(\int^{\Lambda}_{L^{-1}}d^dq
{1\over{q}^2})}<\psi^+_{\sigma}(\vec{r})\psi_{\sigma}(\vec{r})>, 
\end{equation}
\end{mathletters}
which vanishes in the $e\rightarrow\infty$ limit, indicating that the 
bare and chargeless fermions have zero wavefunction overlap as expected. 
Using Eq.\ (\ref{o1}), it is straightforward to show that the 
bare-fermion occupation number $n(\vec{k})$ has no discontinuity across 
fermi surface in the $e^2\rightarrow\infty$ limit (3D), even if the 
chargeless fermions form a fermi liquid, in agreement with the marginal 
fermi liquid picture.   

  The bosonization result we obtained in previous section suggests 
that in the limit $e^2\rightarrow\infty$ the low energy physics of our 
system may be described as a fermi liquid of the chargeless fermions. 
To see whether this is the case we consider the equation of motion for 
the chargeless fermion operators $\psi_{P\sigma}$, assuming that they 
can be treated as canonical fermions. It is straightforward to show that
\begin{eqnarray}
\label{eqqs}
{\partial\psi_{P\sigma}(\vec{r})\over\partial\tau} & = & {1\over2m}\left(
\nabla^2\psi_{P\sigma}(\vec{r})-[\nabla^2\phi(\vec{r})-
:(\nabla\phi(\vec{r}))^2:]\psi_{P\sigma}(\vec{r})-
2\nabla\phi(\vec{r}).\nabla\psi_{P\sigma}(\vec{r})\right)
\\   \nonumber
& & -{1\over{L}^d}\sum_{\vec{q}}v(q)\rho_{ph}(\vec{q})e^{i\vec{q}.\vec{r}}
\psi_{P\sigma}(\vec{r})-{i\over{L}^{d/2}}\sum_{\vec{q}\lambda}
e^{i\vec{q}.\vec{r}}{e\over{m}}\vec{A}_{ph}(\vec{q},{\lambda}).
\nabla\psi_{P\sigma}(\vec{r}),
\end{eqnarray}
where we have neglected constant energy terms coming from
normal ordering of operators inderiving Eq.\ (\ref{eqqs}). Notice that 
direct couplings between fermions and collective modes are absent
in the equation of motion of $\psi_{P\sigma}(\vec{r})$. However, 
interactions between chargeless fermions and particle-hole excitations 
remain in the equation of motion. Moreover, an indirect coupling to 
collective modes is also generated from the fermion kinetic energy term 
as is in the similar "small polaron" problem\cite{mahan}. It is easy 
to see by direct power counting of $e^2$ in the $\phi(\vec{r})$
field that the coupling of the "dressed" fermion to collective 
excitations through the kinetic energy term is much weaker than the 
original fermion-plasmon and fermion-photon couplings. In particular,
the self-energy correction of chargeless fermions from $\phi(\vec{r})$ 
fields remains finite in the $e^2\rightarrow\infty$ limit.

   We shall now study the dynamics of the chargeless fermions in more 
detail. First we write down an effective action for the chargeless 
fermion operator $\psi_{P\sigma}(\vec{r})$, the effective action is 
constructed with the requirement that it reproduces the equation of 
motion \ (\ref{eqqs}) for $\psi_{P\sigma}$. It is straightforward to 
show that the correct action is
\begin{mathletters}
\label{seff}
\begin{equation}
S_{eff}(\psi_P,\psi_P^+)=\sum_{\sigma}\int_0^{\beta}d\tau{d}^dx
\psi^+_{P\sigma}(\vec{x},\tau)\left[
{\partial\over\partial\tau}-{\nabla^2\over2m}-\mu+i\vec{A}
_{eff}(\vec{x},\tau).\nabla+\phi_{eff}(\vec{x},\tau)\right]
\psi_{P\sigma}(\vec{x},\tau),
\end{equation}
where $\phi_{eff}(\vec{q})\sim{v}(\vec{q})\rho_{ph}(\vec{q})$ and
$\vec{A}_{eff}(\vec{q})\sim{e}\vec{A}_{ph}(\vec{q})/m$, with
dynamics given by
\begin{eqnarray}
S_{eff}(\phi) & = & -{1\over2}
\sum_{\vec{q},i\omega}\chi_0(\vec{q},i\omega)|\phi_{eff}(\vec{q},i\omega)|^2,
\\  \nonumber
S_{eff}(\vec{A}) & = & -{1\over2}\sum_{\vec{q},i\omega}
\bar{\chi}_t(\vec{q},i\omega)\vec{A}_{eff}
(\vec{q},\omega).\vec{A}_{eff}(-\vec{q},-i\omega)
\end{eqnarray}
\end{mathletters}
where $\bar{\chi}_t(\vec{q},i\omega)=\chi_t(\vec{q},i\omega)+mn_0$ is 
the {\em total} transverse current susceptibility of free fermions. 
We have neglected in $S_{eff}(\psi_P,\psi_P^+)$ the collective mode
contributions to the dynamics of $\psi_{P\sigma}$ through the kinetic 
energy term. The residue plasmon-fermion interaction induces a weak 
short-ranged effective attractive interaction between chargeless fermions 
and may lead to superconductivity. We shall not discuss this possibility 
in this paper. 

  Fermions interacting with gauge fields with effective interaction
\ (\ref{seff}) have been studied in detail by a number of 
authors\cite{t2,hal,gauge,khm} and it is believed that interaction
between fermions and transverse gauge field with effective 
dynamics \ (\ref{seff}) may lead to break down of fermi liquid theory
and the formation of a marginal fermi liquid. If this is the case, 
the ground state formed by the chargeless fermions would itself be in a 
marginal fermi liquid state and our system can be viewed as a "double" 
marginal fermi liquid state of the original fermions. This is in contrast 
to (Gaussian) bosonization theory which predicts that the chargeless
fermions should form a {\em fermi liquid} state. The failure of 
Gaussian theory in producing a marginal fermi liquid state shows the
limitation of Gaussian theory. This point will be elaborated further
in the last section.    

    It is interesting to make comparison between our approach and the
usual treatment of the $U(1)$ gauge field in t-J model where the gauge 
field effectively imposes the constraints $\rho_s(q)+\rho_h(q)=0$ and
$\vec{j}_s(\vec{q})+\vec{j}_h(\vec{q})=0$\cite{t1}, where $\rho_s
(\vec{j}_s)$ and $\rho_h(\vec{j}_h)$ are the spin and hole densities
(current densities) operators, respectively. In the usual 
treatment\cite{t1,ln}, 
the spins and holes were integrated out at Gaussian level, resulting in 
an effective action for gauge fields very similar to Eq. \ (\ref{seff}) 
except that the longitudinal and transverse response functions
$\chi_0(q,\omega)$ and $\bar{\chi}_t(q,\omega)$ are replaced respectively 
by the sum of the longitudinal and transverse response functions from
the spin- and hole- components, respectively. It is then assumed that 
the spinons and holons in the projected Hilbert space can be treated as 
free particles interacting with the gauge fields with dynamics given by
$S_{eff}$\cite{t1}. Although our analysis here includes only fermions and 
is not directly applicable to the t-J model, the form of our low 
energy effective Lagrangian is in agreement with this assumption, 
suggesting that the spinons in the usual treatment of t-J model are
strongly related to our chargeless fermions. A detailed treatment of
the t-J model using our approach will be reported in a different paper. 

   Finally we examine the validity of replacing the fermion density 
operator $\rho(\vec{r})$ by the average fermion density $n_0$ in the 
diamagnetic term when we derive the equation of motions for the bare 
and chargeless fermions. We shall now show that this approximation is 
justified in the $e^2\rightarrow\infty$ limit for the chargeless 
fermions $\psi_{P\sigma}$. We look at the correction term in the 
equation of motion for chargeless fermions,
$[\Delta{H},\psi_{P\sigma}(\vec{r})]$, where
\[
\Delta{H}={e^2\over2m}
\int{d}^dr\vec{A}(\vec{r}).\vec{A}(\vec{r})(\rho(\vec{r})-n_0). \]
It is easy to show from Eqs.\ (\ref{rho1}), \ (\ref{cles1}) and
\ (\ref{qu2}) that $[\rho(\vec{q}),\psi_{P\sigma}(\vec{r})]
\sim{1\over{e}^2}\psi_{P\sigma}$, $e\vec{A}/m\sim{e}^{1/2}
(\gamma_{\lambda}+\gamma^+_{\lambda})$ and $[e\vec{A}(\vec{r}),\psi_{P\sigma}
(\vec{r})]\sim\nabla\psi_{P\sigma}$ in the $e^2\rightarrow\infty$ limit. 
As a result we have
\[
[\Delta{H},\psi_{P\sigma}(\vec{r})]\sim(\vec{A}(\vec{r}').\nabla
\psi_{P\sigma}(\vec{r}))(\rho(\vec{r})-n_0)+O({1\over{e}}).  \]
The first term vanishes in the $e^2\rightarrow\infty$ limit if the 
matrix elements $<n|(\rho(\vec{r})-n_0)|m>$ vanish faster than
$1/e^{1/2}$ in the physical Hilbert space of interests. This will
be the case if the physical Hilbert space is spanned by the chargeless
fermion operators we construct, because of the commutation rule between 
density operator and chargeless fermions (Eq.\ (\ref{cles1})). 

\section{summary}
   Using a bosonization approach we studied in this paper a gas of 
spin $S=1/2$ fermions interacting with a $U(1)$ gauge field with high
momentum cutoff $q<\Lambda<<k_F$. In particular we consider the
$e^2\rightarrow\infty$ limit where the gauge field becomes confining.
We have analysed the problem in two steps. Within the Gaussian 
approximation we show in section III that a liquid state solution is 
obtained where the particle-hole excitation spectrum of the system is 
always fermi-liquid like, with the charge carried by the particle-hole 
excitation vanishing continuously in the $e^2\rightarrow\infty$ limit. 
To verify the above picture we construct in section IV $S=1/2$, chargeless
fermionic operators that can be used as the starting point for 
constructing quasi-particles in the system. We find that the dynamics
of these "physical" single-particle operators are governed by a
Lagrangian very similar to the effective Lagrangain obtained in
conventional treatment of gauge field model\cite{t1}. The solution we 
obtained describes a kind of 'double-marginal' fermi liquid state where 
(i)the chargeless fermions have vanishing wavefunction overlap with 
bare fermions in the system and (ii)they form a marginal fermi liquid
state themselves. 

  It is useful to distinguish the effects of longitudinal and
transverse gauge fields on our system, in particular how they contribute
in the $e^2\rightarrow\infty$ limit, to the construction of chargeless 
fermions and their effective dynamics. First we consider the longitudinal 
gauge field. The main effect of longitudinal gauge field appears 
through fermion-plasmon interaction which is singular in the limit $e^2\rightarrow\infty$. In dimension $d=2$, there exists an additional 
infra-red singularity associated with the response of plasmon field
(orthogonality catastrophe effect) when charged particles are added to the
system\cite{wen}. The effect exists for any finite values of $e^2>0$. 
The interaction between fermions and particle-hole excitations
is finite and regular for any value of $e^2$. The singular fermion-plasmon
interaction and the infra-red singlarity at dimension $d=2$ are eliminated
together with our canonical transformation \ (\ref{qu1})\cite{ng1},
and the remaining interaction between the chargeless fermions and 
longitudinal gauge fields is regular and introduces no further
singlarities in the chargless fermion system\cite{ng1}.

   The situation is however, quite different with the transverse
gauge field. There exists no infra-red singularity in the 
fermion-photon interaction at both two- and three- dimensions when 
$e^2$ is finite, and the singularity associated with the fermion-photon
interaction in the limit $e^2\rightarrow\infty$ is in fact weaker than 
the corresponding singularity associated with fermion-plasmon 
interaction, as can be seen from direct power counting of $e^2$ in 
$\phi_l$ and $\vec{W}$ fields. As a result, the canonical 
transformation \ (\ref{qu1}) is dominated by effect of plasmons 
in the limit $e^2\rightarrow\infty$. On the other hand, the interaction 
between bare fermions and particle-hole excitations through
transverse gauge field is infra-red singular for any values of $e^2$ 
in both two- and three- dimenions\cite{t2}, and the singularity carries 
over to the effective interaction \ (\ref{seff}) between dressed 
fermions and transverse gauge field. It is believed that the 
infra-red singularity in interaction between fermions and 
particle-hole excitations through transverse gauge field will lead to 
the formation of a marginal fermi liquid state\cite{t2,hal,gauge,khm}.

  The possibility of formation of an marginal fermi liquid state of 
the chargeless fermions points out the limitation of bosonization
theory in Gaussian approximation, where a fermi-liquid type particle-hole
excitation spectrum is always obtained in the limit $e^2\rightarrow\infty$.
It is clear that higher-order terms in cumulant expansion of transverse 
gauge field has to be included in bosonization theory to obtain the 
correct marginal fermi liquid behavior\cite{t2}, and a complete
theory where particle-hole and single-particle excitations are treated
self-consistently is still missing.

  An immediate question that comes with this observation is that we have 
used the excitation spectrum in Gaussian approximation to construct 
chargeless fermion operators in our paper. If the particle-hole 
excitation spectrum is incorrect, our results based on chargeless 
fermion operators $\psi_{P\sigma}$ become questionable! Fortunately 
although the particle-hole excitation spectrum in Gaussian
theory is questionable, the plasmon and photon excitations we obtained 
are exact in the $q\rightarrow0$ or $e^2\rightarrow\infty$ limits. In
particular our canonical transformation \ (\ref{qu1}) for the chargeless 
fermions $\psi_{P\sigma}$ involves only the collective mode operators and 
is independent of the Gaussian approximation! However, it has to be
cautioned that in deriving the effective action $S_{eff}(\psi_P,\psi_P^+)$  
which describes scattering of chargeless fermions with particle-hole 
excitations, it is assumed that the eigen-particle-hole excitations are
described correctly by Gaussian theory. Strong modifications may occur if
the spectrum of particle-hole excitation is modified strongly in the
correct theory. 
 
  Another more fundamental question is how valid it to treat our 
chargeless fermions as canonical fermions, given that they obey a rather 
non-trival commutation relation. In particular, it is possible that the
unusual commutation relation between fermions may lead to additional
non-fermi liquid behaviour. At present we have no answer to these
questions. 

  This work is supported by HKUGC through RGC grant HKUST6124/98P.

\end{document}